\documentclass[times,twocolumn,final]{elsarticle}

\usepackage{medima}
\usepackage{framed,multirow}

\usepackage{amssymb}
\usepackage{latexsym}

\usepackage{url}

\usepackage{hyperref}
\usepackage[table,xcdraw]{xcolor}
\usepackage{multirow}
\usepackage[ruled,linesnumbered]{algorithm2e}

\definecolor{newcolor}{rgb}{.8,.349,.1}

\journal{Medical Image Analysis}

\begin{document}

\verso{Given-name Surname \textit{et~al.}}

\begin{frontmatter}

\title{Cross-Modal Vertical Federated Learning for MRI Reconstruction\tnoteref{tnote1}}%
\tnotetext[tnote1]{Yunlu Yan and Hong Wang contributed equally to this work.}

\author[1]{Yunlu \snm{Yan}}
\author[2]{Hong \snm{Wang}}
\author[2]{Yawen \snm{Huang}}
\author[2]{Nanjun \snm{He}}
\author[1,4]{Lei \snm{Zhu}}
\author[2]{Yuexiang \snm{Li}\corref{cor1}}
\cortext[cor1]{Corresponding author: Yuexiang Li, Yong Xu \textit{email:} vicyxli@tencent.com; yongxu@ymail.com.}
\author[3]{Yong  \snm{Xu}\corref{cor1}}
\author[2]{Yefeng \snm{Zheng}}

\address[1]{ROAS Thrust, The Hong Kong University of Science and Technology (Guangzhou), Guangzhou, China}

\address[2]{Tencent Jarvis Lab, Shenzhen, China}

\address[3]{Shenzhen Key Laboratory of Visual Object Detection and Recognition, Harbin Institute of Technology (Shenzhen), Shenzhen,  China.}

\address[4]{Department of Electonic and Computer Engineering, The Hong Kong University of Science and Technology, Hong Kong SAR, China}


\begin{abstract}
Federated learning enables multiple hospitals to cooperatively learn a shared model without privacy disclosure. Existing methods often take a common assumption that the data from different hospitals have the same modalities. However, such a setting is difficult to fully satisfy in practical applications, since the imaging guidelines may be different between hospitals, which makes the number of individuals with the same set of modalities limited. To this end, we formulate this practical-yet-challenging cross-modal vertical federated learning task, in which {\itshape data from multiple hospitals have different modalities with a small amount of multi-modality data collected from the same individuals}. To tackle such a situation, we develop a novel framework, namely Federated Consistent Regularization constrained Feature Disentanglement (Fed-CRFD), for boosting MRI reconstruction by effectively exploring the overlapping samples (\emph{i.e.,} individuals with multi-modalities) and solving the domain shift problem caused by different modalities. Particularly, our Fed-CRFD involves an intra-client feature disentangle scheme to decouple data into modality-invariant and modality-specific features, where the modality-invariant features are leveraged to mitigate the domain shift problem. In addition, a cross-client latent representation consistency constraint is proposed specifically for the overlapping samples to further align the modality-invariant features extracted from different modalities. Hence, our method can fully exploit the multi-source data from hospitals while alleviating the domain shift problem. Extensive experiments on two typical MRI datasets demonstrate that our network clearly outperforms state-of-the-art MRI reconstruction methods.  The source code will be publicly released upon the publication of this work.

\end{abstract}

\begin{keyword}
\MSC 41A05\sep 41A10\sep 65D05\sep 65D17

\KWD Cross-Modal\sep Domain Shift\sep MRI Reconstruction \sep Vertical Federated Learning
\end{keyword}

\end{frontmatter}


\section{Introduction}\label{sec:intro}
Magnetic resonance imaging (MRI) is one of the most important diagnostic tools in real-world clinical applications. Nevertheless, due to the complex imaging process, the acquisition time of MRI is much longer than other techniques, such as computed tomography (CT). To accelerate the data acquisition, various efforts have been made for the high-quality MRI reconstruction with the under-sampled $k$-space measurements. 
During the past few years, deep learning-based approaches~\cite{wang2016accelerating,sriram2020end, sriram2020grappanet,feng2021accelerated} have been the dominant way for MRI reconstruction and achieved the impressive image reconstruction performance. However, these approaches need a large-scale dataset for centralized training. It is difficult for a single institution to collect a large amount of training data, due to the expensive acquisition cost. Gathering the data from multiple institutions to increase the amount of training data may be a potential solution. However, limited by the data privacy, it is infeasible in realistic scenario.

\begin{figure}[!t]
    \centering
    \includegraphics[width=\linewidth]{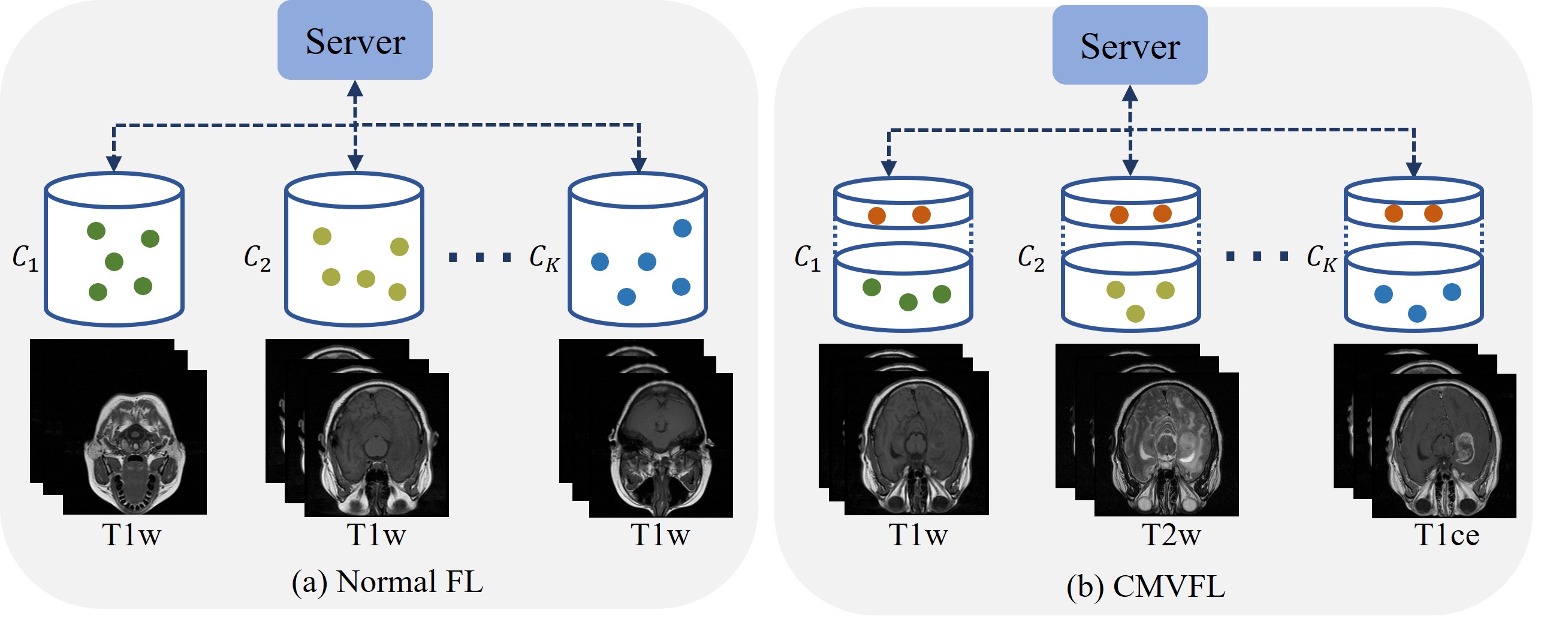}
    \caption{ Illustration of different federate learning task settings, including (a) normal FL and (b) our proposed CMVFL, where the circles within each client represents individuals. Specifically, in (a), every client $C_{k}$ has the same modality data which is hard to satisfy in real scenarios. In contrast, in (b), our clients can execute the MRI reconstruction based on different modalities. Besides, the data from the overlapping individuals (marked by red circles) among different clients can be fully exploited to address the domain shift problem.} \label{image}
    \vspace{-3mm}
    \label{fig:intro}
\end{figure}

Federated learning (FL), an attractive distributed learning paradigm, which allows multiple institutions to cooperate on the premise of protecting privacy \cite{yang2019federated,9084352,ZHANG2021106775}, has been adopted to solve the aforementioned problems \cite{guo2021multi}. The previous attempts were based on the same assumption that the data from multiple hospitals shared the same modalities, as shown in Fig.~\ref{fig:intro}~(a). 
However, in the realistic practice, due to the expensive costs of imaging and the different roles of modalities, each hospital may only acquire one or some of the modalities for specific diagnosis. For example, in low and middle income countries, the patients may first be screened in a community hospital via T1 modality and then transferred to
the more specialized hospital, where additional MRI modalities will be scanned as supplements for an exact diagnosis (\emph{e.g.,} T2). 
Thus, there may be some individuals with different modalities in the community and specialized hospitals, 
\emph{i.e.,} {\itshape overlapping samples} in this study, as shown in Fig.~\ref{fig:intro}~(b). In contrast, the healthy or severe patients may only have the single modality in the community and specialized hospitals, respectively.
Besides, the loss of data and the alteration of imaging instruments may also lead to the `missing modalities' problem.

In this paper, we first formulate a practical-yet-challenging federated learning task, termed cross-modal vertical federated learning (CMVFL), where the clients have different modalities of different patients but with a small set of overlapping samples, as shown in Fig.~\ref{fig:intro}~(b). Under such a setting, the existing FL methods easily suffer from the domain shift problem, which is caused by different modalities of clients.
To this end,
we propose a vertical federated learning framework, namely federated consistent regularization constrained feature disentanglement (\textbf{Fed-CRFD}). Specifically, we introduce a novel intra-client feature disentanglement scheme to separate the modality-invariant features from the modality-specific ones.
Furthermore, compared to the traditional FL problem, the key of CMVFL is to effectively utilize the vertical samples (\emph{i.e.,} overlapping samples) to improve the performance of federated learning. 
Therefore, to effectively exploit the consistent information of vertical subject in different clients, we propose a cross-client feature alignment module for the modality-invariant features in the latent space, which can finally lead to a global consistent representation for MRI reconstruction. 
Extensive experiments demonstrate that our method can effectively exploit the information within overlapping samples and achieve the better MRI reconstruction performance than the existing approaches.
Our contributions can be summarized as follows:
\begin{itemize}
    \item We formulate a practical cross-modal vertical federated learning scenario, which aims to achieve the better reconstruction performance by exploiting the useful knowledge from vertical/overlapping samples.
    \item We propose an intra-client feature disentanglement scheme to obtain the consistent feature distribution, which can effectively address the domain shift problem. Moreover, a cross-client latent representation consistency scheme for vertical samples is also proposed to mitigate the bias among different clients, which further significantly improves the MRI reconstruction performance.
    \item We conduct extensive experiments on a publicly available fastMRI dataset and a private clinical dataset for the MRI reconstruction task, which shows the superiority of our method, \emph{i.e.,} a new state-of-the-art is achieved.
\end{itemize}

\section{Related Works}
In this section, we briefly review the related works including federated learning, representation disentanglement learning, and MRI reconstruction.

\subsection {Federated Learning} FedAvg~\cite{mcmahan2017communication}, the most popular method, introduces a trusted server to average the updated model parameters from the local clients and further updates global model, which works well when the data is independent and identically distributed (IID). However, it has been demonstrated that FedAvg suffers from the client-drift issue caused by statistical heterogeneity ~\cite{li2019convergence,karimireddy2020scaffold}. The client-drift leads to the inconsistent local objectives, which influences the average procedure and then results in a suboptimal global model. Besides, the different imaging protocols adopted in different institutions would lead to the domain shift, which is a type of statistical heterogeneity. Against this issue, Guo \emph{et al.} \cite{guo2021multi} constructed the first FL framework called FL-MRCM for MRI reconstruction, which tries to solve the domain-shift problem by aligning the feature distribution of the source sites and the target sites. Feng \emph{et al.} \cite{feng2021specificity} proposed another strategy to handle the domain shift problem by preserving client heterogeneity. Although the previous methods have achieved some promising performance improvement, there are still some problems that have not been taken into consideration. For example, one key point is that these methods are proposed based on the task setting that different institutions have the same modal data. However, in practice, different institutions may use different modalities due to the clinical diagnostic needs. Besides, these methods aim to tackle the horizontal FL problem based on the assumption that the data among different clients is non-overlapping. As seen, they do not fully explore the information of the overlapping samples. 


\subsection {Representation Disentanglement Learning} Representation disentanglement learning, a typical solution to domain shift, aims to extract a general representation from multi-domain data ~\cite{peng2019domain, liu2018detach} or multi-modal data ~\cite{ouyang2021representation, fei2021deep}. The effectiveness of such disentanglement representations has been widely substantiated in various applications, such as myocardial segmentation~\cite{chartsias2018factorised}, multi-modal brain tumor segmentation~\cite{chen2019robust}, multi-modal deformable registration~\cite{qin2019unsupervised} and domain adaptation~\cite{9832940,9558836}. However, these methods rely on the centralized dataset, and do not fully explore the potential of feature disentanglement for medical imaging based on datasets from multiple hospitals. For instance, Chen \emph{et al.} \cite{chen2019robust} introduced a gating fusion strategy to fuse different modality-invariant codes for learning modality-invariant representations, which is hard to execute in distributed clients. Moreover, the methods ~\cite{liu2018detach, fei2021deep} using the same model to directly trains on multi-modal or multi-domain data will divulge privacy that is not allowed by the rule of FL.

\subsection {MRI Reconstruction} MRI reconstruction aims to reconstruct the high-quality MR images based on the under-sampled $k$-space measurements, which makes it possible to accelerate the MR imaging. 
Against the ill-posed inverse problem, traditional methods~\cite{haldar2010compressed,he2016accelerated} proposed to adopt the prior knowledge to regularize the solution space. 
With the rapid development of deep learning, many MRI reconstruction methods have been proposed in the past few years~\cite{feng2021T2Net, yang2018admm, wang2016accelerating}. Attributed to the powerful representation capability of convolutional neural network (CNN) and the large  training datasets, such deep learning based methods generally perform better than the traditional optimization based approaches. To reduce the cost of paired data (\emph{i.e.,} low-quality and high-quality images) collection and alleviate the data dependence, unsupervised learning based MRI reconstruction methods have been designed~\cite{cole2020unsupervised, korkmaz2022unsupervised}. However, their performance is generally not comparable to that of the fully-supervised methods, therefore cannot satisfy the need of real applications. Against this issue, in this paper, we propose to adopt the distributed training paradigm for MRI reconstruction where multiple clients can cooperatively train a reconstruction model without data sharing. In this manner, the data among different clients can be sufficiently utilized, which indirectly reduces the requirement of collecting a large amount of paired data and has the potential to achieve better reconstruction performance than single client.

\section{Problem Statement}
In this section, we formulate the paradigm of cross-modal vertical federated learning task, and introduce the basic federated learning framework for MRI reconstruction in details.


\vspace{2mm}
\subsubsection{Cross-Modal Vertical Federated Learning} 
\begin{figure*}[!t]
    \centering
    \includegraphics[width=\textwidth]{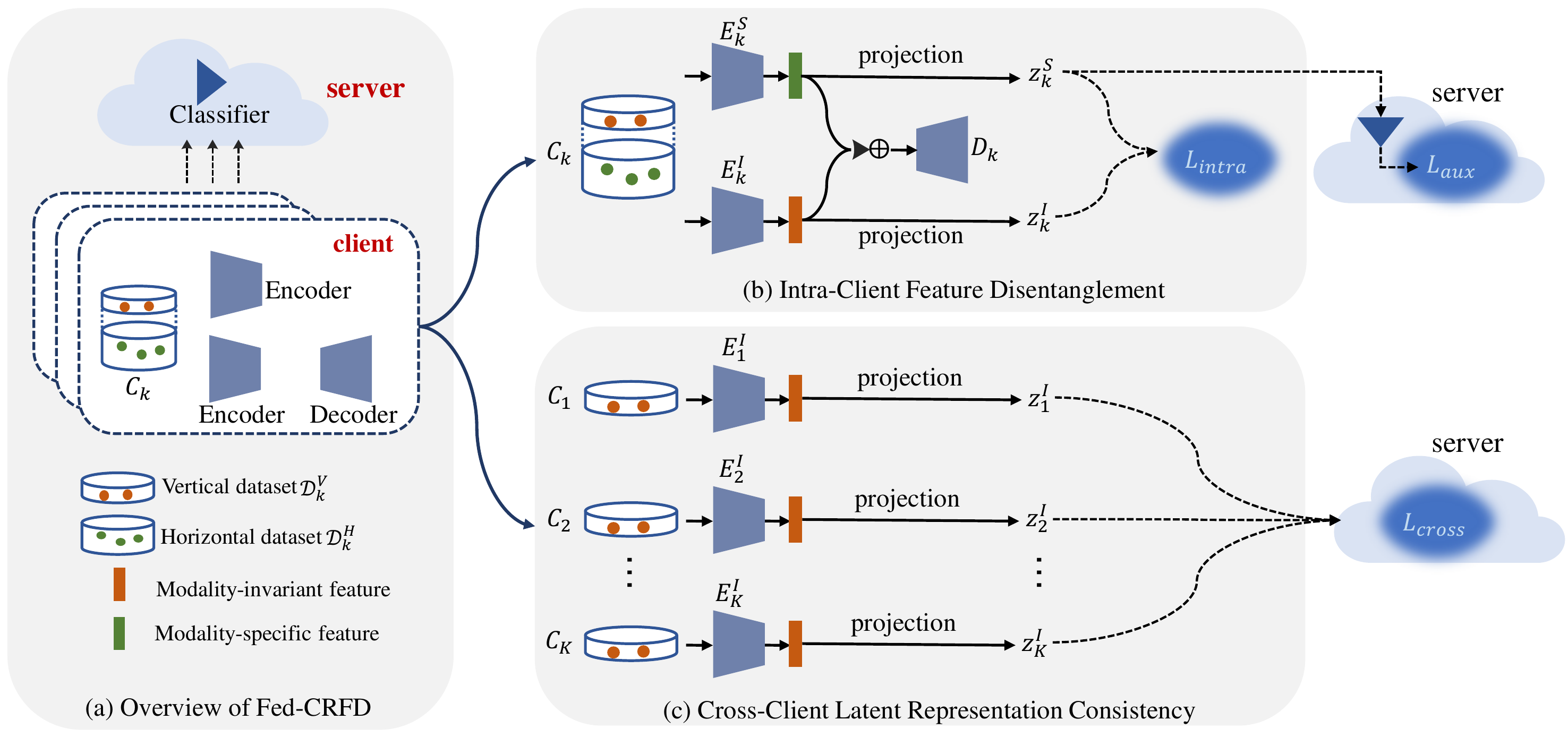}
    \caption{(a) The overview of the proposed Fed-CRFD framework; (b) The intra-client feature disentanglement module; and (c) cross-client latent representation consistency module.} \label{fig1}
\end{figure*}

As shown in Fig.~\ref{fig1}, in the cross-modal vertical federated learning task, there are $K$ clients denoted as $\{C_{1}, \ldots, C_{K}\}$. For each client $C_{k}$, it is composed of the specific MR image modality $\mathcal{M}_{k,j}$ and contains the horizontal dataset $\mathcal{D}^{H}_{k}$ with ${n}^{H}_{k}$ samples and the vertical dataset $\mathcal{D}^{V}_{k}$ with ${n}^{V}_{k}$ samples, where $\mathcal{M}_{k,j} \in \mathcal{M}$, and $\mathcal{M}=\{\mathcal{M}_{1}, \mathcal{M}_{2},  \ldots, \mathcal{M}_{J}\}$. $\mathcal{M}$ represents the set of the commonly-adopted MR modalities, such as $\{{T1w}, {T2w}, {T1ce}\}$ shown in Fig.~\ref{fig:intro}~(b). Clearly, for client $C_{k}$, its private dataset $\mathcal{D}_{k}$ is $\mathcal{D}^{H}_{k} \cup \mathcal{D}^{V}_{k}$ and the total number of samples is $n_{k}={n}^{H}_{k}+{n}^{V}_{k}$.


\vspace{1.5mm}
\noindent{\bf Horizontal Dataset:} In our settings (Fig.~\ref{fig:intro}~(b)), each client separately captures the specific modal MR images of different patients. $\mathcal{D}^{H}_{k}$ represents the MR images from the non-overlapping subjects among different clients and it is exclusive to the client $C_{k}$. 


\vspace{1.5mm}
\noindent{\bf Vertical Dataset:} $\mathcal{D}^{V}_{k}$ consists of the MR images from the overlapping subjects among different subjects. Hence, ${n}^{V}_{1} = {n}^{V}_{2}=\cdots= {n}^{V}_{K}$. All the vertical dataset from $K$ clients constitute the entire vertical space $\mathcal{D}^{V}$, \emph{i.e.}, $\mathcal{D}^{V}=\mathcal{D}^{V}_{1}\cup \mathcal{D}^{V}_{2} \cup \cdots \cup \mathcal{D}^{V}_{K} $. Clearly, $\mathcal{D}^{V}$ contains the MR images of the overlapping subjects with multi-modalities as $\{\mathcal{M}_{1,j}, \mathcal{M}_{2,j}, \ldots, \mathcal{M}_{K,j}\}$.


It should be noted that the overlapping subjects in $\mathcal{D}^{V}$ can be identified via encrypted entity alignment~\cite{nock2018entity} without privacy disclosure. Thus, consistent to \cite{kang2020fedmvt,cheng2021secureboost}, the vertical/overlapping subjects (\emph{i.e.,} different modalities of the same individual) from different clients can be identified by the server. 


\vspace{2mm}
\subsubsection{Basic FL for MRI Reconstruction} The goal of FL is to train a globally optimal MRI reconstruction model in a distributed manner by minimizing the empirical loss of each client. The global objective function of standard FL paradigm~\cite{mcmahan2017communication} can be written as:
\begin{equation}
    \mathcal{L} = \sum_{k=1}^{K} p_{k} \mathcal{L}_{k}, \quad {with} \    p_{k}=\frac{n_{k}}{\sum_{i=1}^{K}n_{i}}, \label{eq:1}
\end{equation}
where $\mathcal{L}_{k}$ is a local objective function of the $k$-th client $C_k$ and $p_{k}$ is the weighting coefficient.


For MRI reconstruction, current deep learning-based methods aim to utilize convolutional neural networks to learn the mapping function $f$ from an under-sampled image $x$ to the high-quality image $y$. Correspondingly,  the local objective function for the client $C_{k}$ can be formulated as:
\begin{equation}
    \mathcal{L}_{k} = \mathbb{E}_{(x, y) \sim \mathcal{D}_{k}} \left\| f(W_{k};x)-y \right\|_{1},\label{eq:2}
\end{equation}
where $\| \cdot \|_{1}$ denotes the $l_{1}$ loss; $(x,y)$ represents a paired data from $\mathcal{D}_{k}$; and $W_{k}$ denotes the local model parameters of $C_{k}$. The global model parameters $W_{G}$ are updated by averaging the local model parameters~\cite{mcmahan2017communication}:
\begin{equation}
    W_{G} = \sum_{k=1}^{K} p_{k} W_{k}.\label{eq:3}
\end{equation}

Directly applying such basic FL framework for MRI reconstruction would suffer from the domain shift caused by different modalities. Against this issue, we fully explore the horizontal dataset $\mathcal{D}^{H}_{k}$ and vertical dataset $\mathcal{D}^{V}_{k}$, and specifically propose a network framework for the MRI reconstruction task, which will be presented in the next section.

\section{The Proposed Fed-CRFD Framework}
\label{sec:method}
In this section, the proposed federated consistent regularization constrained feature disentanglement framework, called Fed-CRFD, is introduced in details. As shown in Fig.~\ref{fig1}, the pipeline of Fed-CRFD consists of two components, \emph{i.e.,} intra-client feature disentanglement module (b) and cross-client latent representation consistency module (c). These two modules aim to address the domain shift problem existing in different modalities and exploit the useful information contained in overlapping samples from different clients to better help feature extraction, respectively.

\vspace{1.5mm}
\noindent {\bf Local Model:} Concretely, each local model in our Fed-CRFD consists of two encoders (modal-specific ${E}^{S}$ and modality-invariant ${E}^{I}$) and one decoder ${D}$. The two encoders are implemented for feature disentanglement, and the well-trained local model ($\{{E}^{S}, {E}^{I}\} \cup {D}$) is used for MRI reconstruction in practical applications.


\subsection{Intra-Client Feature Disentanglement}
The domain shift between different modalities leads to an unaligned feature distribution, which degrades the performance of the global model. To tackle this problem, we propose an intra-client feature disentanglement framework to decompose the representations of a subject into {\itshape modality-specific} and {\itshape modality-invariant} features, respectively, as shown in Fig.~\ref{fig1} (b). Specifically,

\subsubsection {Modality-Specific Representations} 
At the client $C_{k}$, a modal-specific encoder ${E}^{S}_{k}$ is deployed to extract modality-specific representations via an auxiliary classification task. To be specific, for an image $x\in\mathcal{D}_{k}$, the encoder ${E}^{S}_{k}$ extracts the feature and then projects it to a latent space as:
\begin{equation}
    z^{S}_{k} = \mathcal{P}\left({E}^{S}_{k}(x)\right),
\label{eq:4}
\end{equation}
where $\mathcal{P}(\cdot)$ denotes the projection operation from  image feature to a latent space, achieved by global average pooling and flattening, and $z^{S}_{k}$ is the learnt latent feature. Then, a multi-layer perceptron (MLP) classifier $\mathcal{C}$ is implemented at the server to conduct the auxiliary classification task, \emph{i.e.,} identifying the modality of the latent features from different clients. Under such a supervision, the encoder ${E}^{S}_{k}$ is encouraged to extract only modality-specific information, which is beneficial for feature disentanglement. The objective function of auxiliary classification task can be defined as:
\begin{equation}
    \mathcal{L}_{aux} = CE(\mathcal{C}(z^{S}_{k}), V_{k}),
\label{eq:5}
\end{equation}
where $CE(\cdot)$ is the cross entropy loss function and $V_{k}$ is the one-hot modality label.~\footnote{For simplicity, we omit the index $k$ in $\mathcal{L}_{aux}$, as well as $\mathcal{L}_{intra}$ and $\mathcal{L}_{recon}$ presented later.} Although $z^{S}_{k}$ and $V_{k}$ are transferred from client to server, the features in latent space are irreversible~\cite{guo2021multi,wu2021federated}. Meanwhile, the one-hot modality label is data-independent, and thus contains no related information of patients. Therefore, our approach does not violate the protocol of privacy protection.

\vspace{2mm}
\subsubsection {Modality-Invariant Representations} Apart from modal-specific encoder ${E}^{S}_{k}$, an encoder of reconstruction network ${E}^{I}_{k}$ is proposed to encode the same input of ${E}^{S}_{k}$. The encoder ${E}^{I}_{k}$ is enforced to learn the modality-invariant representations $z^{I}_{k}$. To achieve that, we minimize the similarity between $z^{I}_{k}$ and $z^{S}_{k}$ to distill the modality-specific features. The objective function can be formulated as:
\begin{equation}
    \mathcal{L}_{intra} = - \|z^{I}_{k} - z^{S}_{k}\|_{1}, \ {with} \ z^{I}_{k} = \mathcal{P}\left({E}^{I}_{k}(x)\right).\label{eq:6}
\end{equation}

\vspace{2mm}
\subsubsection {MRI Reconstruction} The two encoders ${E}^{S}_{k}$ and ${E}^{I}_{k}$ can learn complementary feature representations with the proposed feature disentanglement constraints. Therefore, for the better reconstruction performance, we fuse the features via element-wise summation and feed it into the decoder ${D}_{k}$ to obtain the reconstruction result. The objective of the reconstruction task is formulated as:
\begin{equation}
    \mathcal{L}_{recon} = \|{D}_{k}({E}^{I}_{k}(x) + {E}^{S}_{k}(x) ) - y \|_{1}.\label{eq:7}
\end{equation}
The local model learns the modality-invariant representations, which are independent of the hospitals; therefore, the feature distribution of the learnt model is unified. In this regard, we average the parameters of local models to obtain the global model $W_{G}$:
\begin{equation}
    W_{G} = \sum_{k=1}^{K} p_{k} W_{k}, \quad {with} \ W_{k} = W_{{E}^{I}_{k}} \cup W_{{D}_{k}},\label{eq:8}
\end{equation}
where $W_{{E}^{I}_{k}}$ and $W_{{D}_{k}}$ are the parameters of the encoder ${E}^{I}_{k}$ and the decoder ${D}_{k}$, respectively.

\subsection{Cross-Client Latent Representation Consistency}
Although the unified feature distributions can be achieved using the proposed intra-client feature disentanglement,
due to the lack of constraints between different clients during local model updating, the feature representations for the vertical samples may be disparate across clients. To address this problem, we propose a cross-client latent representation consistency scheme as shown in Fig. \ref{fig1} (c), which effectively leverages the useful knowledge from the overlapping samples. Concretely, we transfer the modality-invariant latent features from each client to the server, and force them to be consistent by maximizing the similarity between each pair from the feature pool. The objective function on the vertical space $D^{V}_{k}$ among different clients is formulated as:
\begin{equation}
    \mathcal{L}_{cross} = \sum_{i=1, i\neq k}^{K} \|z^{I}_{k} - z^{I}_{i}\|_{1}.
    \label{eq:9}
\end{equation}
The loss calculation is executed on the server, which then returns the gradients to clients for local model updating. In general, the cross-client consistency is beneficial for the feature alignment among all clients. 

\begin{algorithm}[!t]
\caption{Training Fed-CRFD for MRI Reconstruction}

\small
\label{alg:cmvfl}
\KwIn{$K$ clients with datasets $(\mathcal{D}_{1}^{H}, \mathcal{D}_{1}^{V}), \ldots$, $(\mathcal{D}_{K}^{H}, \mathcal{D}_{K}^{V})$, batchsize $N_{b}$, communication rounds $T$, local epochs $Q$, learning rate $\eta$, classifier $\mathcal{C}$, modality-specific encoders (${E}^{S}_{1}, {E}^{S}_{2}, \ldots, {E}^{S}_{K}$),
modality-invariant encoders (${E}^{I}_{1}, {E}^{I}_{2}, \ldots, {E}^{I}_{K}$), decoders (${D}_{1}, {D}_{2}, \ldots, {D}_{K}$)} 
\KwOut{Global model $W^{(T)}_{G}$}

\textbf{Global Aggregation:}\\
Initialize $W^{(0)}_{G}$, $W^{(0)}_\mathcal{C}$, $W^{(0)}_{{E}_{1}^{S}}, W^{(0)}_{{E}_{2}^{S}}, \ldots, W^{(0)}_{{E}_{K}^{S}}$\\
\For{round $t=1,2,...,T$}
{
    \For{client $k=1,2,...,K$ \textbf{parallelly}}{
        Receive the global model $W_{G}^{(t-1)}$ from server\;
        
        $W_{k}^{(t)}$ $\leftarrow$ $\textbf{Local Training}\left(k, W_{G}^{(t-1)}\right)$ \;
    }
    $W^{(t)}_{G}$ $\leftarrow$ $\sum_{k=1}^{K} p_{k}W_{k}^{(t)}$\;
}
Return $W^{(T)}_{G}$\\ 
\textbf{Local Training}$\left(k, W_{G}^{(t-1)}\right)$:\\
        $W_{k}^{(t)}$ $\leftarrow$ $W_{G}^{(t-1)}$\;

        \For{epoch $q=1,2,...,Q$}{
        
            \For{horizontal dataset $\mathcal{D}_{k}^{H}$}{
             $(x,y) \leftarrow $SampleMiniBatch($\mathcal{D}_{k}^{H}$, $N_{b}$)\\
                $\mathcal{L}=\mathcal{L}_{recon} + \mu_{1} \mathcal{L}_{aux} + \mu_{2}\mathcal{L}_{intra}$\;
                $\left(W_{k}^{(t)}, W^{(t)}_{{E}_{k}^{S}}, W^{(t)}_{\mathcal{C}}\right)$  $\!\!\!\leftarrow\!\!\!$ $\left(W_{k}^{(t)}, W^{(t)}_{{E}_{k}^{S}}, W^{(t)}_{\mathcal{C}} \right) - \eta \nabla \mathcal{L}$\;
            }
            
            \For{vertical dataset $\mathcal{D}_{k}^{V}$}{
                $(x,y) \leftarrow $SampleMiniBatch($\mathcal{D}_{k}^{V}$, $N_{b}$)\\
                $\mathcal{L}=\mathcal{L}_{recon} + \mu_{1} \mathcal{L}_{aux} + \mu_{2}\mathcal{L}_{intra} + \mu_{3} \mathcal{L}_{cross} $\\
             $\left(W_{k}^{(t)}, W^{(t)}_{{E}_{k}^{S}}, W^{(t)}_{\mathcal{C}}\right)\!\!\!\leftarrow$ \quad \quad \quad \quad $\left(W_{k}^{(t)}, W^{(t)}_{{E}_{k}^{S}}, W^{(t)}_{\mathcal{C}} \right) - \eta \nabla \mathcal{L}$\;
            }
            
        }
Return $W_{k}^{(t)}$ to the server.
\end{algorithm}

\subsection{Updating Procedure}
We provide the detailed training procedure of the proposed Fed-CRFD for MRI reconstruction in Alg.~\ref{alg:cmvfl}. In each communication round, the client $C_{k}$ receives the parameters of the latest global model $W_{G}^{(t-1)}$ from the server, and updates the local model (${E}^{I}_{k}$ and ${D}_{k}$). For the horizontal space $D^{H}_{k}$, all models will be updated via:
\begin{equation}
    \mathcal{L} = \mathcal{L}_{recon} + \mu_{1} \mathcal{L}_{aux} + \mu_{2}\mathcal{L}_{intra},
    \label{eq:10}
\end{equation}
where $\mu_{1}$ and $\mu_{2}$ are the loss weights. While for the vertical space $D^{V}_{k}$, with our cross-client latent representation consistency, the overall objective $\mathcal{L}$ for the local model can be formulated as:
\begin{equation}
    \mathcal{L} = \mathcal{L}_{recon} + \mu_{1} \mathcal{L}_{aux} + \mu_{2}\mathcal{L}_{intra} + \mu_{3} \mathcal{L}_{cross},
    \label{eq:11}
\end{equation}
where $\mu_{3}$ is the relative weight of $\mathcal{L}_{cross}$.

\section{Experiments}
In this section, we present the experimental results and conduct a comprehensive analysis.

\subsection{Experimental Setup}
\subsubsection{Datasets} We evaluate our method on the public \textbf{fastMRI}\footnote{https://fastmri.org/dataset/}~\cite{zbontar2018fastmri} dataset and a private clinical dataset. The modalities of T1w and T2w from the fastMRI dataset are adopted in experiments, which contain 352 volumes in total. Each volume has 16 slices. The size of each slice is $320 \times 320$ pixels. Besides, the private dataset is desensitized and collected from our collaborative hospital. The dataset has 126 volumes with both T1w and T2w modalities while each volume contains 20$\sim$62 slices. The original resolution of slice ranges from $208\times 256$ pixels to $512 \times 512$ pixels.
Hence, we resize the slices to $200 \times 200$ pixels. We randomly divide each dataset to training and test sets by the ratio of 80:20. The data partition is patient-wise, \emph{i.e.,} the two sets involve no overlapping patients.



\vspace{1mm}
\subsubsection {Data Distribution Setting}
    To simulate the CMVFL setting, we randomly select some patients in the training set as the vertical space, and partition the two modalities of the same patient into different clients. The proportion of vertical samples are controlled by parameter $\beta$. Then, the rest patients in the training set is assigned to the clients by selecting one of the modalities without overlapping. Thus, the clients have either T1w modality or T2w modality, respectively. To conform the different imaging protocols of hospitals in practice, we adopt different under-sampling 
    protocols and accelerating
 rates: \textit{1D Uniform 5$\times$} and \textit{2D Random 3$\times$}
 ~\cite{wang2016accelerating}.

\begin{table*}[!t]
    \centering
    \caption{Quantitative results on the fastMRI dataset~\cite{zbontar2018fastmri} and private dataset  with different proportions of the vertical samples. We run three trials and report the mean and standard derivation for each method. To compare Fed-CRFD with other methods statistically, we perform paired t-test on PSNR and report the p-value. ($\uparrow$ means the higher the better)}\label{tab1}    
    \small
    \setlength{\tabcolsep}{3mm}{
    {\begin{tabular}{l|c|c|c|c|c|c}
    \hline
    \multirow{2}{*}{Method}  & 
    \multicolumn{3}{c|}{$\beta$ = 10\%} & \multicolumn{3}{c}{$\beta$ = 2\%}\\
    \cline{2-7}
     & PSNR$\uparrow$ & SSIM$\uparrow$  & p-value& PSNR$\uparrow$ & SSIM$\uparrow$  & p-value \\
    \hline\hline
    \multicolumn{7}{c}{\textbf{Dataset 1: fastMRI}} \\
    \hline\hline
    Solo & 31.06 $\pm$ 0.15 & 0.8759 $\pm$ 0.0029  &0.0037 & 30.84 $\pm$ 0.15 & 0.8707 $\pm$ 0.0024 & 0.0013 \\
    Centralized (Upper bound) & 37.30 $\pm$ 0.17 & 0.9537 $\pm$ 0.0021 & - & 37.13 $\pm$ 0.13 & 0.9517 $\pm$ 0.0022  &-\\
    \hline
    FedAvg \cite{mcmahan2017communication} & 33.30 $\pm$ 0.21 & 0.9315 $\pm$ 0.0013    & 0.0206 & 33.15 $\pm$ 0.27 & 0.9304 $\pm$ 0.0017  & 0.0297 \\
    FL-MRCM \cite{guo2021multi} & 33.37 $\pm$ 0.12 & 0.9301 $\pm$ 0.0012 & 0.0015 & 33.24 $\pm$ 0.13 & 0.9283 $\pm$  0.0008 & 0.0182\\
    
    FedMRI \cite{feng2021specificity} & 34.15 $\pm$ 0.31 & 0.9328 $\pm$ 0.0033 & 0.0098 & 34.03 $\pm$ 0.27 & 0.9309 $\pm$  0.0024 & 0.0142\\
    
    \hline
    \rowcolor[HTML]{EFEFEF}  Fed-CRFD (Ours) & \textbf{35.35 $\pm$ 0.22} & \textbf{0.9372 $\pm$ 0.0022}  & - & \textbf{35.19 $\pm$ 0.24} & \textbf{0.9358 $\pm$ 0.0036}  & -\\
    \hline\hline
    \multicolumn{7}{c}{\textbf{Dataset 2: Private Dataset}} \\
    \hline\hline
     Solo & 29.10 $\pm$ 0.19 & 0.7759 $\pm$ 0.0032  &0.0040 & 28.92 $\pm$ 0.16 & 0.7715 $\pm$ 0.0033 & 0.0039 \\
    Centralized (Upper bound) & 31.11 $\pm$ 0.09 & 0.8593 $\pm$ 0.0011 & - & 31.02 $\pm$ 0.11 & 0.8571 $\pm$ 0.0016  &-\\
    \hline
    FedAvg \cite{mcmahan2017communication} & 29.95 $\pm$ 0.15 & 0.7933 $\pm$ 0.0010    & 0.0197 & 29.86 $\pm$ 0.17 & 0.7912 $\pm$ 0.0008  & 0.0102 \\
    FL-MRCM \cite{guo2021multi} & 30.04 $\pm$ 0.05 & 0.7926 $\pm$ 0.0006 & 0.0069 & 29.95 $\pm$ 0.03 & 0.7917 $\pm$ 0.0006 & 0.0008\\
    
    FedMRI \cite{feng2021specificity} & 30.32 $\pm$ 0.20 & 0.8149 $\pm$ 0.0047 & 0.0235 & 30.13 $\pm$ 0.17 & 0.8075 $\pm$  0.0036 & 0.0177\\
    
    \hline
    \rowcolor[HTML]{EFEFEF}  Fed-CRFD (Ours) & \textbf{30.95 $\pm$ 0.07} & \textbf{0.8391 $\pm$ 0.0008}  & - & \textbf{30.76 $\pm$ 0.06} & \textbf{0.8373 $\pm$ 0.0010}  & -\\
     \hline
    \end{tabular}}
    }
\end{table*}

\begin{table*}[t]
    \centering
    \caption{Results on the two different datasets with three modalities and three clients. ($\uparrow$ means the higher the better)}\label{tab3}
    \small
    \setlength{\tabcolsep}{2.0mm}{
    {\begin{tabular}{l|c|c|c|c|c|c}
    \hline
    \multirow{2}{*}{Method}  &  \multicolumn{3}{c|}{fastMRI~\cite{zbontar2018fastmri}} & \multicolumn{3}{c}{Private Dataset}\\
    \cline{2-7}
     & PSNR$\uparrow$ & SSIM$\uparrow$  & p-value& PSNR$\uparrow$ & SSIM$\uparrow$  & p-value \\
    \hline\hline
    Solo & 31.58 $\pm$ 0.39 & 0.8162 $\pm$ 0.0105  &0.0072 & 28.32 $\pm$ 0.35 & 0.6572 $\pm$ 0.0108 & 0.0141 \\
    Centralized (Upper bound) & 36.12 $\pm$ 0.28  & 0.9406 $\pm$ 0.0049 & - & 30.52 $\pm$ 0.18 & 0.8349 $\pm$ 0.0048  &-\\
    \hline
    FedAvg \cite{mcmahan2017communication} & 33.77 $\pm$ 0.15 & 0.9192 $\pm$ 0.0010    & 0.0273 & 30.24 $\pm$ 0.07 & 0.8041 $\pm$ 0.0008  & 0.0098 \\
    FL-MRCM \cite{guo2021multi} & 33.74 $\pm$ 0.05 & 0.9265 $\pm$ 0.0006 & 0.0067 & 30.05 $\pm$ 0.17 & 0.7977 $\pm$ 0.0076 & 0.0011\\
    
    FedMRI \cite{feng2021specificity} & 34.56 $\pm$ 0.32 & 0.9326 $\pm$ 0.0031 & 0.0083 & 30.37 $\pm$ 0.25 & 0.8265 $\pm$  0.0081 & 0.0162\\
    
    \hline
    \rowcolor[HTML]{EFEFEF}  Fed-CRFD (Ours) & \textbf{35.65 $\pm$ 0.18} & \textbf{0.9389 $\pm$ 0.0019}  & - & \textbf{30.86 $\pm$ 0.14} & \textbf{0.8358 $\pm$ 0.0030}  & -\\
    \hline
    \end{tabular}}}
\end{table*}

\vspace{1mm}
\subsubsection {Implementation Details}
We employ U-Net~\cite{ronneberger2015u}, which is officially released by fastMRI,\footnote{https://github.com/facebookresearch/fastMRI/} as the reconstruction model. The model is optimized by the Adam optimizer and the learning rate is 0.0001. The framework is implemented with PyTorch on a single NVIDIA RTX 3090 GPU with 24GB of memory and the batch size $N_{b}$ is 16. The $\mu_{1}$, $\mu_{2}$ and $\mu_{3}$ are empirically set to 0.01, 0.01, and 0.01, respectively. In addition, we train a total of 50 communication rounds with 2 local epochs per round. For a fair comparison, we train all comparing methods with the same protocol and all methods are observed to converge by the end of training procedure.

\begin{figure*}[!t]
    \centering
    \includegraphics[width=\textwidth]{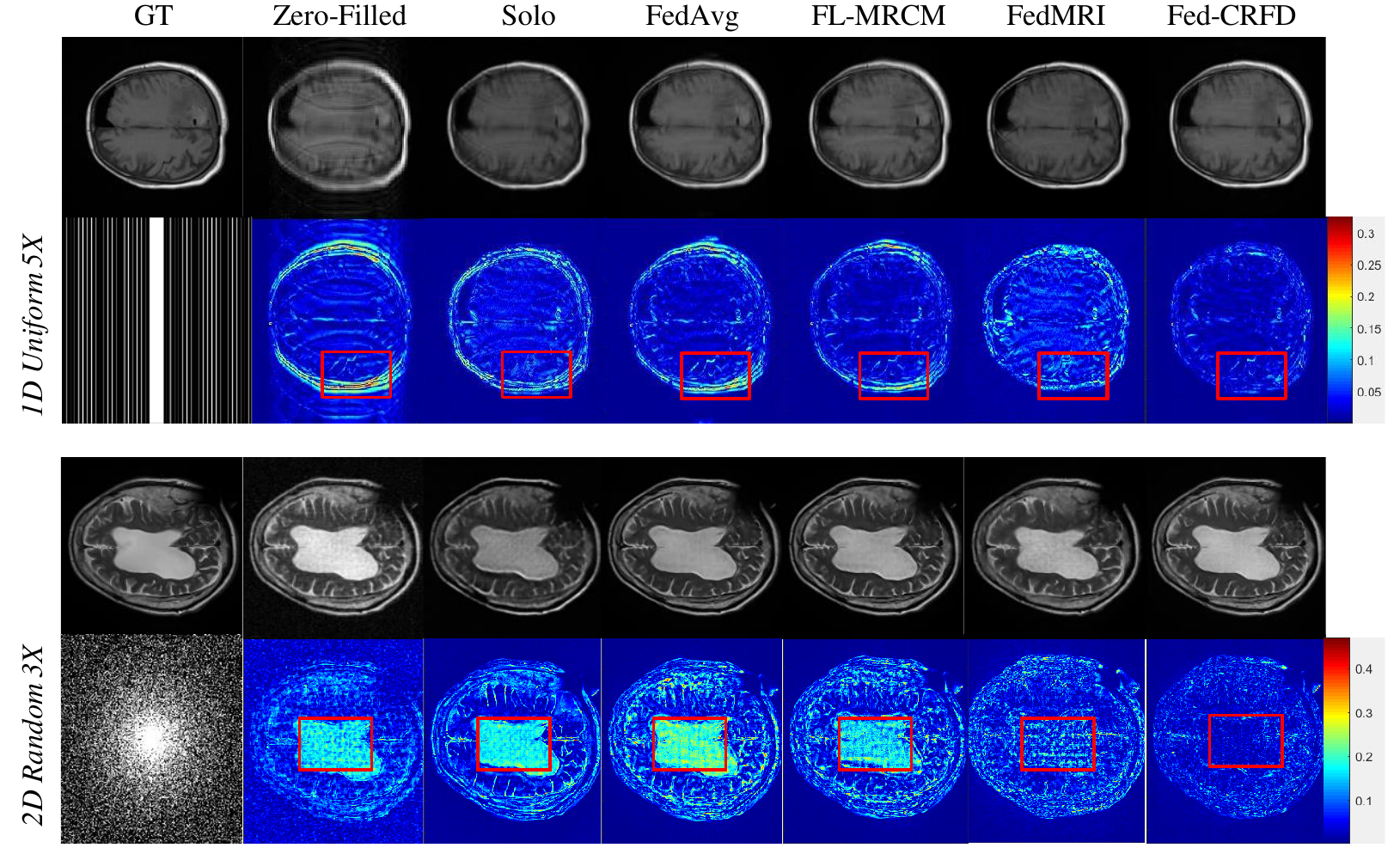}
    \caption{Reconstructed MRI with different methods on the fastMRI~\cite{zbontar2018fastmri} dataset.} \label{fastmri_errormap}
\end{figure*}
\begin{figure*}[!t]
    \centering
    \includegraphics[width=1\textwidth]{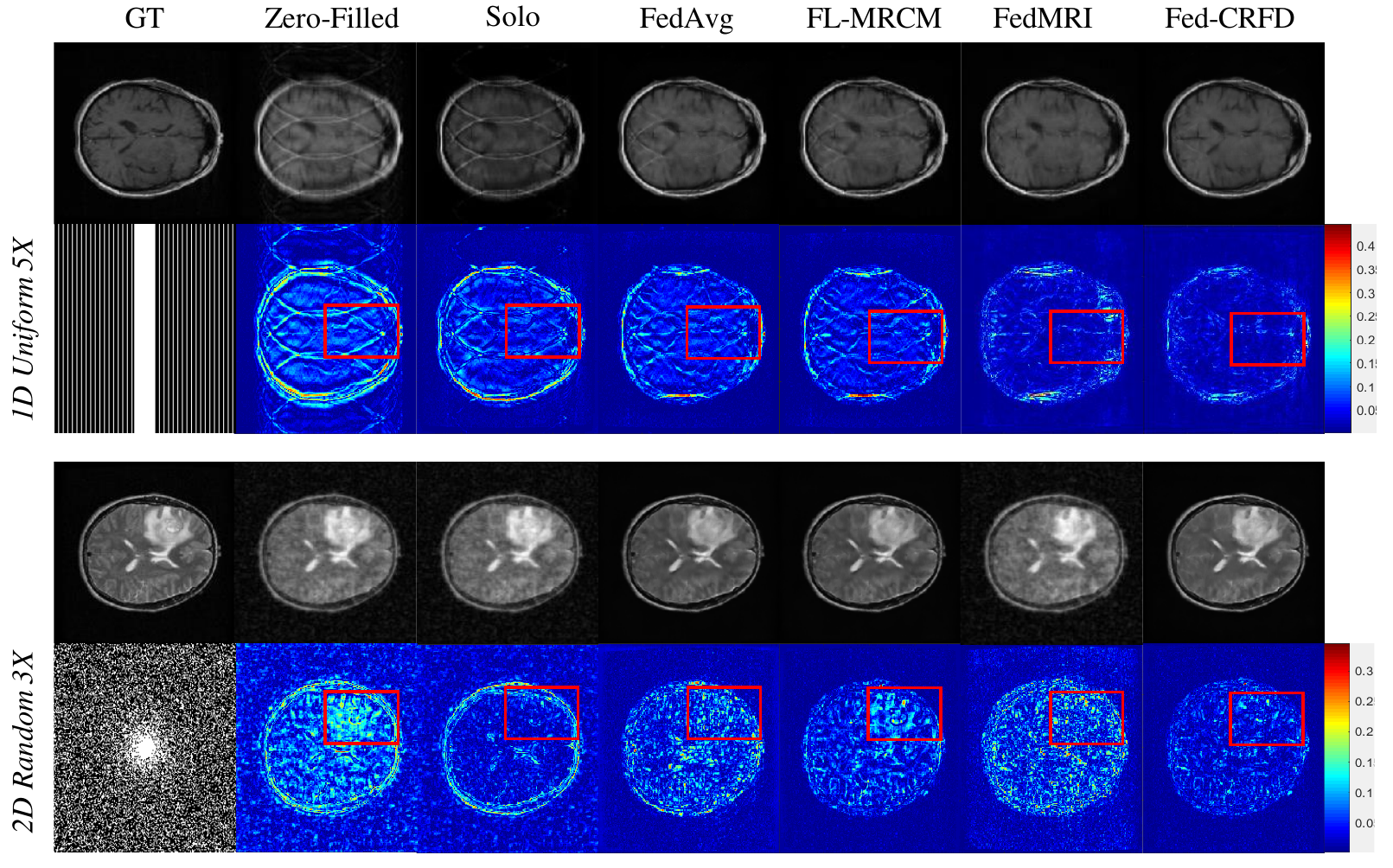}
    \caption{Reconstructed MRI with different methods on the Private dataset.} \label{tiantan_errormap}
    \vspace{-2mm}
\end{figure*}

\subsection{Performance Evaluation}
To validate the effectiveness of our method, we involve the following baselines for comparison:

\begin{itemize}
    \setlength{\itemsep}{0pt}
    \setlength{\parsep}{-2pt}
    \setlength{\parskip}{-0pt}
    \setlength{\leftmargin}{-15pt}
    \item \textbf{Solo}: The client trains model on its private data without communication.
    \vspace{1mm}
    \item \textbf{Centralized}: We gather all local data of the clients for centralized training, which is the theoretical upper bound of FL.
    \vspace{1mm}
    \item \textbf{FedAvg} \cite{mcmahan2017communication}: It is the most commonly used baseline, which directly averages all local model parameters after a communication round. 
    
    \vspace{1mm}
    \item \textbf{FL-MRCM} \cite{guo2021multi}: This is one of the state-of-art FL method for MRI Reconstruction, which aligns the feature distributions of the source sites to the target site, and we implement this framework using the public code.\footnote{https://github.com/guopengf/FL-MRCM} 
    
    \vspace{1mm}
    \textcolor{black}{
    \item \textbf{FedMRI} \cite{feng2021specificity}: 
    The method is a personalized FL based framework, which learns the shared information and client-specific properties to tackle the domain shift problem. The publicly available code is adopted for implementation.\footnote{https://github.com/chunmeifeng/FedMRI}}
\end{itemize}

We report the MRI reconstruction performances on two datasets in Table \ref{tab1} under different settings (\emph{i.e.,} $\beta = 10\%$ and $\beta = 2\%$). For FL-MRCM and FedMRI, we adopt the settings of the original papers. To alleviate the influence caused by random initialization, the average PSNR, SSIM scores and standard derivation of three independent trials are calculated for the comparing methods and our Fed-CRFD.  Additionally, to measure the statistical significance of performance improvements, we conduct the paired t-test for Fed-CRFD and each baseline, and report the p-value as well. As seen, the FL methods significantly outperform Solo, which demonstrates the effectiveness of FL in leveraging multiple clients to obtain a robust global model. Compared to FedAvg, the FL-MRCM achieves almost the same PSNR and even lower SSIM scores. Such an experimental result shows that FL-MRCN cannot tackle the domain shift problem from different modalities. Due to learning the specific information of clients, FedMRI can mitigate the modal difference to some extent compared with FedAvg. Besides, the previous horizontal FL frameworks, \emph{i.e.,} FL-MRCM and FedMRI, have not considered the valuable information of vertical samples. In contrast, thanks to the novel design of our method, our method can effectively address the problem, which significantly improves the PSNR scores from $33.30$ dB to $35.35$ dB on the fastMRI dataset and from $29.86$ dB to $30.95$ dB on our private dataset, respectively. Furthermore,  our Fed-CRFD is verified to achieve the significant improvement compared to FL-MRCM and FedMRI, \emph{i.e.,} the corresponding p-value $< 0.05$, and the comparable performance to the centralized model on our private dataset, which show the superiority of our method.

\vspace{1mm}
\subsubsection {Qualitative Comparison} Moreover, to compare the quality of the restored images, we provide some reconstructed MRI images with different methods on the fastMRI dataset and private dataset as shown in Fig.~\ref{fastmri_errormap} and Fig.~\ref{tiantan_errormap}, respectively. The first row contains the ground truth and the restored images and the second row shows the undersampled mask and the corresponding error maps. Less texture in error map indicates better reconstruction performance. Obviously, compared with other methods, our method has the best reconstruction quality, especially for the region is marked by a red box.

\vspace{1mm}
\subsubsection {Results with Fewer Vertical Samples} In practice, compared to horizontal dataset, the vertical dataset is usually very small. To investigate the influence caused by the variations of vertical sample quantity ($\beta$), we evaluate the performance of our method with different numbers of vertical samples and present the results in Table \ref{tab1}. In comparison with the result of $\beta = 10\%$, the performance of all methods degrade due to the reduction in the amount of vertical data when $\beta = 2\%$. For example, the average PSNR score of Fed-CRFD drops from $30.95$ dB to $30.76$ dB and the average PSNR score of FedAvg degrades from  $29.95$ dB to $29.86$ dB on our private dataset, respectively. Nevertheless, the overall results of our method still significantly outperform the other methods, except the centralized approach, which demonstrates the excellent generalization of our method even with the less vertical samples.

\vspace{1mm}
\subsubsection {Results with More Clients} Our Fed-CRFD is a general solution for CMVFL with no limitation on the number of clients. To verify that, in addition to two modalities (\emph{i.e.,} T1w and T2w), we involve the extra T1 post-contrast modality from the fastMRI dataset, and the T1 contrast-enhanced modality from our private dataset, respectively, as a new client in this experiment. Thus, the vertical sample has the three different modalities.
The new client adopts new under-sampling patterns and acceleration rates \textit{1D Cartesian 4$\times$}. The $\beta$ is set to $10\%$ and other experimental settings remain unchanged. We report the experimental results on two datasets in Table \ref{tab3}.

As shown in Table \ref{tab3}, compared to FedAvg, our Fed-CRFD improves the average PSNR score from $33.77$ dB to $35.65$ dB ($+1.88$ dB) and the average SSIM from $0.9192$ to $0.9389$, respectively, on the fastMRI dataset. In comparison, FL-MRCM obtains the lower PSNR and SSIM scores than FedAvg and FedMRI has the better reconstruction results. In addition, the proposed Fed-CRFD still yields a significant improvement compared to FedMRI and FL-MRCM on our private dataset. The experimental results validate the robustness of global generalized feature representation learnt by our Fed-CRFD. Surprisingly, our Fed-CRFD  even outperform the centralized model, which is usually regarded as the upper bound of FL methods.

\begin{table}[!tb]
    \centering
\caption{Quantitative results with more clients on the fastMRI \cite{zbontar2018fastmri} dataset. ($\uparrow$ means the higher the better; UB: Upper bound)}
    \label{tab:mconfastMRI}
    \small
    \setlength{\tabcolsep}{1.4mm}{
    {\begin{tabular}{l|c|c|c}
    \hline
    Method & PSNR$\uparrow$ & SSIM$\uparrow$ & p-value \\
    \hline\hline
    
    Solo & 30.89 $\pm$ 0.44 & 0.7990 $\pm$ 0.0073 & 0.0228 \\
    Centralized (UB) &35.76 $\pm$ 0.27 & 0.9421 $\pm$ 0.0028 & - \\
    \hline
    FedAvg\cite{mcmahan2017communication}  & 32.38 $\pm$ 0.25 & 0.8343 $\pm$ 0.0037  & 0.0031  \\
    FL-MRCM \cite{guo2021multi} & 32.56 $\pm$ 0.52 & 0.8466 $\pm$ 0.0061 & 0.0119 \\
     FedMRI \cite{feng2021specificity} & 33.06 $\pm$ 0.32 & 0.8575 $\pm$ 0.0025 & 0.0257 \\
    \hline
    \rowcolor[HTML]{EFEFEF} 
    Fed-CRFD (Ours) & \textbf{33.82 $\pm$ 0.19} &
    \textbf{0.8625 $\pm$ 0.0033 } &- \\
    \hline
\end{tabular}}}
\vspace{-3mm}
\end{table}

To further explore the scalability of our method, we partition the dataset of each client into two subsets and obtain a larger federation with six clients. Note that there are clients having the same modality in this setting. Particularly, for each client, we downsample the MR images with different under-sampling patterns and acceleration rates, respectively: \textit{1D Uniform 5$\times$}, \textit{2D Random 3$\times$}, \textit{1D Cartesian 4$\times$}, \textit{1D Uniform 3$\times$}, \textit{2D Random 6$\times$}, and \textit{2D Radial 4$\times$}, while other settings are kept the same. We conduct the experiments on the fastMRI dataset with the new setting and the results are reported in Table~\ref{tab:mconfastMRI}. As shown, our Fed-CRFD still achieves the best reconstruction performance compared with other FL methods on the fastMRI dataset, which demonstrates the excellent scalability of our method.

\vspace{1mm}
\subsubsection{Generalization Performance.} We further explore the generalization ability of models trained from three FL methods. Specifically, we train all methods on BraTS dataset~\cite{menze2014multimodal} using only T1w modality and T2w modality. Next, we directly evaluate the model on the test set of fastMRI as shown in Table \ref{gp}. Due to learning the  client specific information that restricts the generalization ability of method, FedMRI has the worse results compared with other methods. Moreover, the result indicates that our method has the best generalization ability compared with the other two FL methods (\emph{i.e.,} FedAvg and FL-MRCM). To further evaluate the generalization of our method, we adopt the reconstructed images for the downstream task. Specifically, we perform the multi-modal (T1w and T2w) segmentation task on our private dataset and the publicly available BraTS dataset \cite{braTS}, respectively, in the next section. 

\begin{table}[!t]
    \centering
    \caption{The result of generalization performance on the fastMRI dataset. ($\uparrow$ means the higher the better)}
    \label{gp}
    \small
    \setlength{\tabcolsep}{3.0mm}{
    {\begin{tabular}{l|c|c}
    \hline
    Method & PSNR$\uparrow$ & SSIM$\uparrow$ \\
    \hline\hline
    FedAvg\cite{mcmahan2017communication}  & 29.51 & 0.6193  \\
    FL-MRCM \cite{guo2021multi} & 29.46 & 0.6853  \\
     FedMRI \cite{feng2021specificity} & 28.59 & 0.5611 \\
    \hline
    \rowcolor[HTML]{EFEFEF} 
    Fed-CRFD (Ours) & \textbf{29.74} & \textbf{0.7000}  \\
    \hline
    \end{tabular}}}
\end{table}

\subsection{Performance on Downstream Task.}
To further evaluate the reconstruction performance of our proposed method, we apply the reconstructed MR images for the downstream segmentation task.

\vspace{1mm}
\subsubsection{Datasets} Specifically, we perform the multi-modal (T1w and T2w) segmentation task on our private dataset and the publicly available BraTS dataset \cite{braTS}, respectively. For the BraTS dataset, we randomly select 70 volumes for training and 30 volumes for test. The full-sampled MR images, which are the ground truth of MRI reconstruction task, are used to train a segmentation model. During the test stage, we simultaneously down-sample two modalities of test set and use the FL-trained reconstruction models to reconstruct the down-sampled MR images. Then the reconstructed images are fed to the trained segmentation model for tumor segmentation. The BraTS dataset provides the official annotations, while our private dataset is annotated by experienced radiologists from a collaborative hospital. Our private datasest is a two-categories segmentation task including background and lesion area. We adopt the Dice score as the performance metric for the quantitative evaluation of the segmentation task. 

\vspace{1mm}
\subsubsection{Implementation Details}  The 3D U-Net is adopted as the segmentation model, which directly concatenates the two modalities as input. For a fair comparison, all FL-methods are trained and evaluated under the same protocol.

\vspace{1mm}
\subsubsection{Performance Evaluation} 
\begin{figure}[!t]
    \centering
    \includegraphics[width=\linewidth]{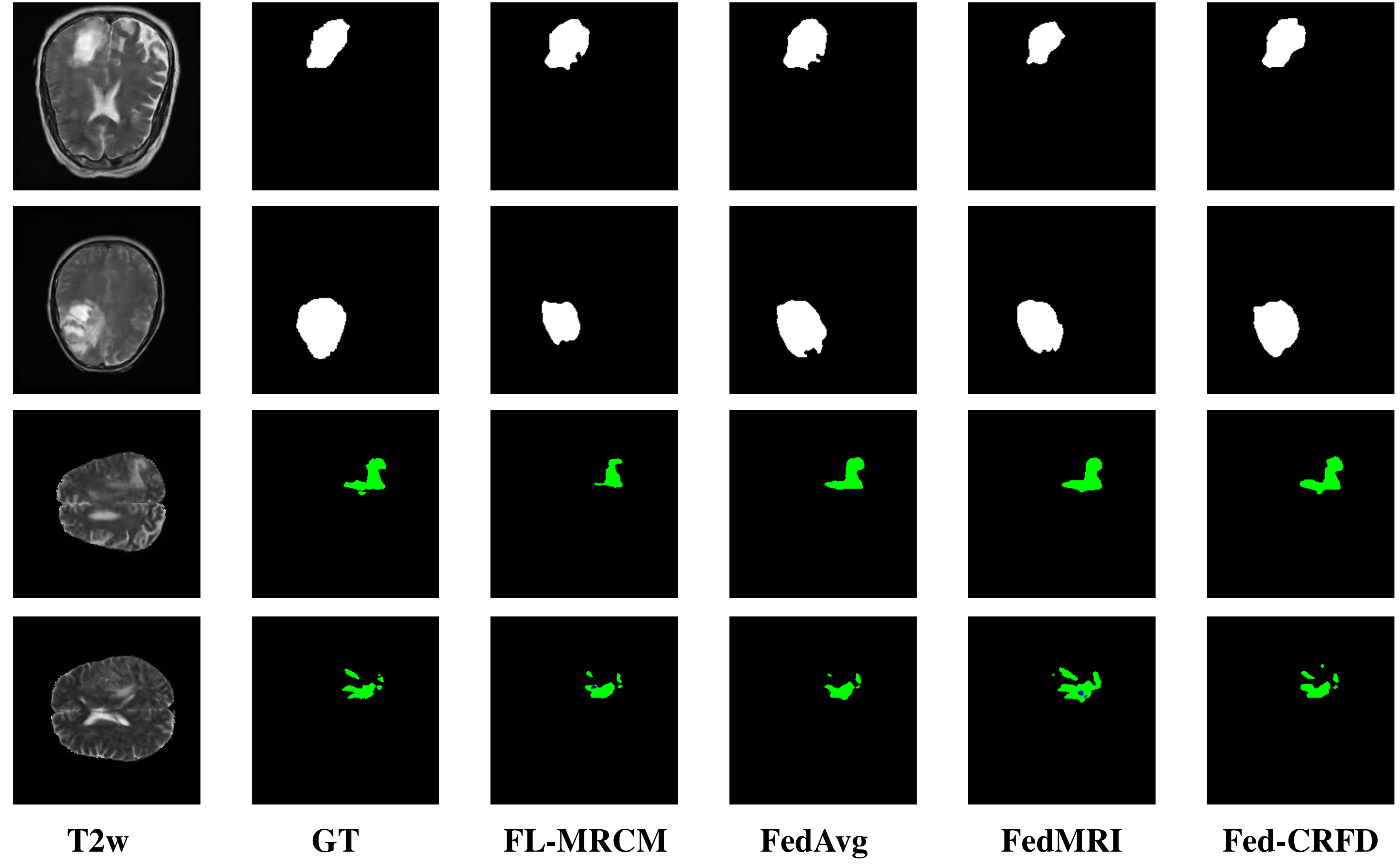}
    \caption{Visual tumor segmentation results on the MR images reconstructed by different methods. Specifically, the top two rows represent the segmentation results on the private dataset and the bottom two rows represent the segmentation result on the BraTS dataset.} \label{tiantan_seg}
\end{figure}

Table \ref{tab5} lists the quantitative results of different FL methods on the downstream segmentation task. As seen, our proposed  Fed-CRFD  consistently achieves higher Dice scores on both Private and BraTS datasets, which finely reflects that our method has the capability to accomplish the better reconstruction results and then boosts the downstream segmentation performance. Fig.~\ref{tiantan_seg} presents the visual results on several MR images. As seen, the proposed Fed-CRFD significantly outperforms the other two methods qualitatively.

\begin{table}[!t]
    \centering
    \caption{Quantitative  results (measured with the Dice coefficient) of the downstream segmentation task on our private dataset and the BraTS ~\cite{menze2014multimodal} dataset.}
    \label{tab5}
    \small
    \setlength{\tabcolsep}{3.0mm}{
    {\begin{tabular}{l|c|c}
    \hline
    Method & Private & BraTS  \\
    \hline\hline
    FedAvg\cite{mcmahan2017communication}  & 78.95\% &51.69\% \\
    FL-MRCM \cite{guo2021multi}  & 79.12\% &50.71\%
    \\
    FedMRI \cite{feng2021specificity} & 80.29\% & 51.61\% \\
    \hline
    \rowcolor[HTML]{EFEFEF} 
    Fed-CRFD (Ours) &\textbf{81.43\%}&\textbf{53.72\%}  \\
    \hline
    \end{tabular}}}
    \vspace{-3mm}
\end{table}

\subsection{Performance Analysis}

\subsubsection {Feature Distributions} 
\begin{figure}[!t]
    \centering
    \includegraphics[width=1.0\linewidth]{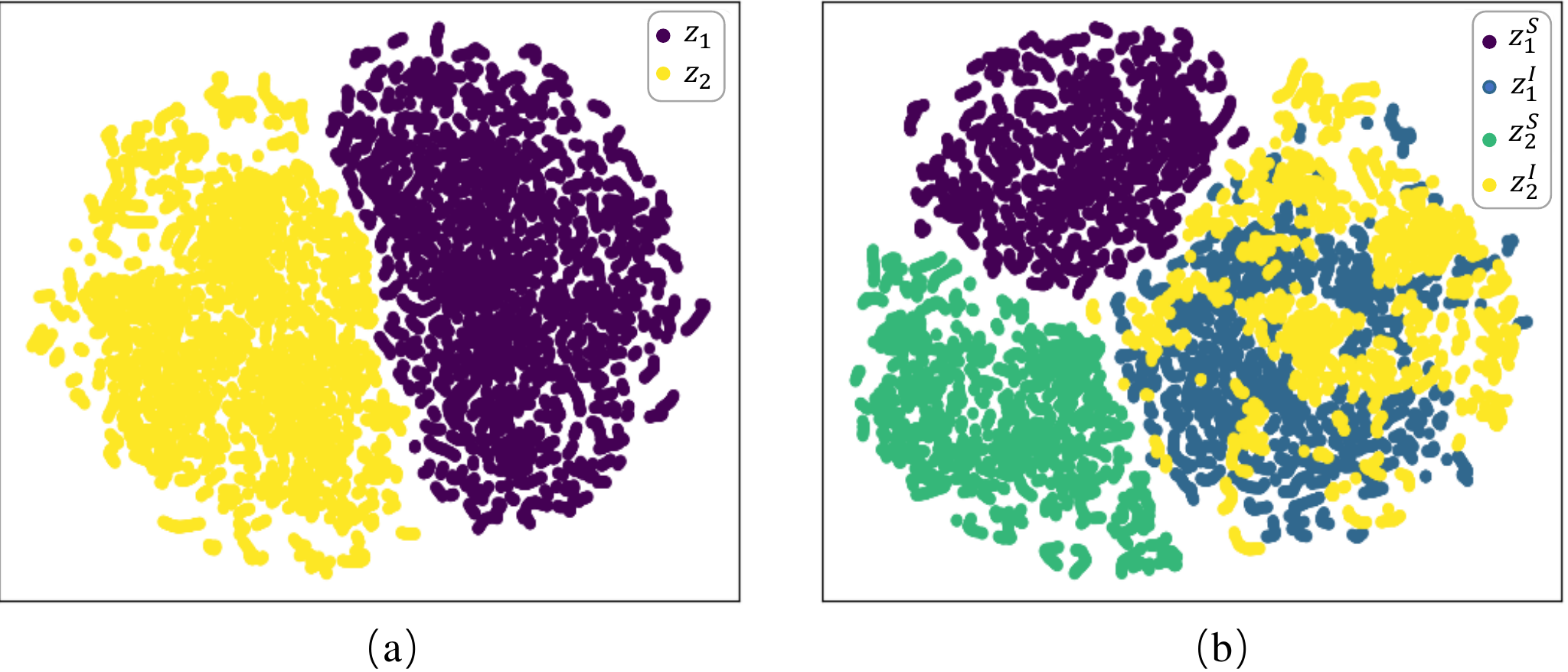}
    \caption{Visualization of the learned latent space with t-SNE on the fastMRI dataset. (a) The original latent feature distributions ($z_{1}, z_{2}$) of two clients. (b) The distributions of latent features learned by our method, where $z^{I}_{*}$ represents the modality-invariant features and the $z^{S}_{*}$ represents the modality-specific features.} \label{tsne}
\end{figure}

To further validate the effectiveness of the mechanism underlying our method, we use the t-SNE~\cite{van2008visualizing} to show the distributions of latent features before and after disentanglement learning in Fig. \ref{tsne}. For the easier visualization and a better understanding of the mechanism, we only use two modalities: T1w and T2w. As shown in Fig. \ref{tsne} (a), there is an obvious distribution gap between two original modalities, which is mainly caused by the domain shift problem, \emph{i.e.,} the modality-specific characteristics. After  disentanglement learning with our framework, we have two observations from Fig. \ref{tsne} (b): 1) modality-specific features are still distinguished from each others; 2) The modality-invariant features are mixed together and form a unified distribution. 
The visualization results reveal the success of feature decoupling achieved by our framework, and the shared reconstruction model learns modality-invariant features from a unified distribution, which can effectively address the domain shift and obtain the better MRI reconstruction performance.

\vspace{1mm}
\subsubsection{Ablation Study} We further investigate the contribution of two key components in our method, \emph{i.e.,} intra-client feature disentanglement and cross-client consistency in latent representations. Since our method degenerates to FedAvg without the two schemes, FedAvg is adopted as the baseline for comparison. Furthermore, the importance of modality-specific features for the reconstruction in Eq. \ref{eq:7} is also evaluated. The ablation study is conducted on the fastMRI dataset and the results are listed in Table \ref{ablationstudy}. It can be observed that the two components are crucial for the success of  our method, \emph{i.e.,} both of them remarkably improve the MRI reconstruction performance. In addition, our method is observed to yield excellent performance even only using the modality-invariant features for reconstruction.

\begin{table}[!t]
    \centering
    \small
    \caption{The result of ablation study on the fastMRI dataset. ($\uparrow$ means the higher the better; consist.: consistency)}    \label{ablationstudy}
    \setlength{\tabcolsep}{3.0mm}{
    {
    \begin{tabular}{l|c|c}
    \hline
     Method & PSNR$\uparrow$ & SSIM$\uparrow$ \\
    \hline\hline
    FedAvg \cite{mcmahan2017communication}  & 33.42 & 0.9332  \\
    Fed-CRFD w/o Cross-client consist.  & 34.57 & 0.9307  \\
    Fed-CRFD w/o Feature addition  & 35.07 & 0.9345   \\
    \hline
    \rowcolor[HTML]{EFEFEF} 
    Fed-CRFD (Ours) & \textbf{35.33} & \textbf{0.9360}  \\
    \hline
    \end{tabular}
    }}
\end{table}

\begin{table}[!t]
    \centering
    \small
    \caption{The reconstruction results of different  similarity measures on the fastMRI dataset. ($\uparrow$ means the higher the better)}    \label{dsm}
    \setlength{\tabcolsep}{2.0mm}{
    {
    \begin{tabular}{l|c|c}
    \hline
     Method & PSNR$\uparrow$ & SSIM$\uparrow$ \\
    \hline\hline

    cosine-distance   & 33.36 & 0.9307  \\
    $l_{2}$-distance  & 33.24 & 0.9287  \\
    \hline
    \rowcolor[HTML]{EFEFEF} 
    Fed-CRFD (Ours) & \textbf{35.33} & \textbf{0.9360}  \\
    \hline
    \end{tabular}
    }}
    \vspace{-3mm}
\end{table}

\vspace{1mm}
\subsubsection{Different  Similarity Measures}
As shown in Eq. \ref{eq:6} and Eq. \ref{eq:9}, we use $l_{1}$ norm distance (\emph{i.e.,} Manhattan distance) to measure the similarity of latent feature representations, which is the key component in our method. In fact, there are other measurements for the similarity between features, \emph{e.g.,} cosine distance and Euclidean distance. To this end, we explore the influence caused by different choices of similarity measurements to our method.
For the sake of comparison, we build up two baselines, denoted as cosine-distance and $l_{2}$-distance, where cosine distance and $l_{2}$ norm distance (\emph{i.e.,} Euclidean distance) are adopted as the similarity measurement term, respectively. We conduct the experiment on the fastMRI dataset with two modalities (T1w and T2w) and $\beta=10\%$. The experimental settings are consistent to the previous ones. For a fair comparison, we finetune the parameters ($\mu_{1}$, $\mu_{2}$ and $\mu_{3}$) of cosine-distance and $l_{2}$-distance and report the best reconstruction results in Table~\ref{dsm}. 
It can be observed that $l_{1}$ distance is significantly superior to other two measurements. The results indicate that $l_{1}$ distance can excellently measure the similarity between different latent feature representations.

\begin{table}[!t]
    \centering
    \small
    \caption{The accuracy (\%) of auxiliary task on the fastMRI dataset.} \label{label:taskacc}
    \setlength{\tabcolsep}{4.0mm}{
    {
    \begin{tabular}{l|c|c}
    \hline
    Modality                    & Training set  & Test set  \\
    \hline\hline
    T1w, T2w                    & 100.00\%      & 100.00\%  \\
    T1w, T2w, T1 post-contrast  & 99.78\%       & 98.53\%   \\
    \hline
    \end{tabular}
    }}
\end{table}

\vspace{1mm}
\subsubsection{Auxiliary Task Accuracy}To decouple the features into modal-specific and modal-invariant components, we formulate an auxiliary modality classification task to separate the modal-specific feature representations (Sec.~\ref{sec:method}) from the modal-invariant ones. Therefore, in this section, we explore the effectiveness of the classifier $\mathcal{C}$ and the modal-specific encoder $E^{S}$ of our method, \emph{i.e.,} whether $\mathcal{C}$ can accurately identify the source modality of feature encoded by $E^{S}$. In Table~\ref{label:taskacc}, we present the accuracy of the classifier on training set and test set of fastMRI dataset with different numbers of modalities, respectively. It is worthwhile to mention that the classifier will only be used in training set for the training of the reconstruction model. The accuracy on test set is evaluated for the validation of its effectiveness.
As shown in Table~\ref{label:taskacc}, the accuracy of auxiliary task yielded by the classifier is satisfactory, \emph{i.e.,} the encoder $E^{S}$ can successfully extract the modal-specific information and embed it into latent space.

\begin{figure}[!t]
    \centering
    \includegraphics[width=\linewidth]{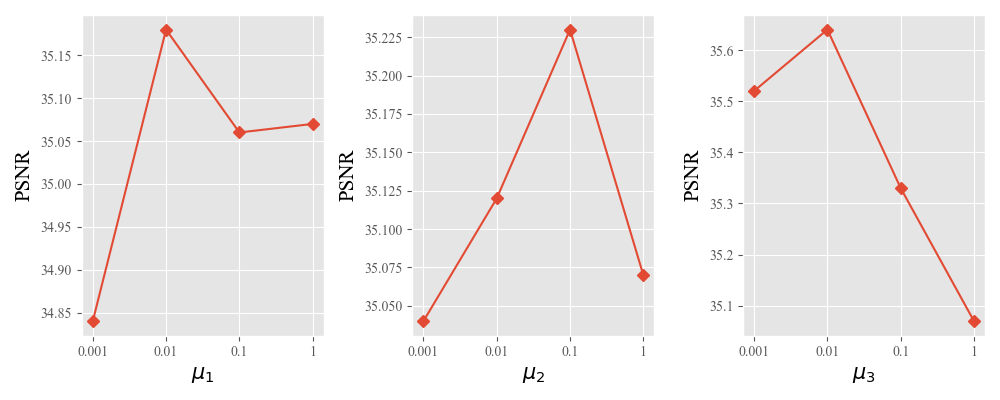}
    \caption{Analysis of three loss weight control hyperparameters, \emph{i.e.,} $\mu_{1}$, $\mu_{2}$, $\mu_{3}$.}\label{image}
    \label{fig:parametr}
\vspace{-4mm}
\end{figure}

\vspace{1mm}
\subsubsection{Hyperparameters Study}The proposed method involves three important hyperparameters, \emph{i.e.,} $\mu_{1}$, $\mu_{2}$, $\mu_{3}$, controlling the three loss functions, \emph{i.e.,} $\mathcal{L}_{aux}$, $\mathcal{L}_{intra}$, $\mathcal{L}_{cross}$, respectively. In this experiment, we evaluate the influence caused by each hyperparameter to the model performance. Specifically, we tune one of the hyperparameters from a set \{0.001, 0.01, 0.1,  1\}, while fixing other two hyperparameters to 1. The experiments are conducted on the fastMRI dataset with two modalities (T1w and T2w). For each experiment, we record the PSNR scores (as shown in Fig.~\ref{fig:parametr}). From the figure, we observe that the reconstruction accuracy is stable ($<$ 0.35 dB) with the variations of hyperparameters ($\mu_{1}$, $\mu_{2}$). The best reconstruction performance is achieved with $\mu_{1}=1$, $\mu_{2}=1$, $\mu_{3}=0.01$. Our Fed-CRFD can yield satisfactory reconstruction performances with the small values of $\mu_{3}$ in a feasible range.

\section{Conclusion}
In the realistic practice, due to the expensive costs of imaging and the different roles of modalities, each hospital may only acquire one or some of the modalities for specific diagnosis. Such a practical problem is not yet explored by researchers. In this work, we formulated this challenging problem as cross-modal vertical federated learning (CMVFL) task that aims to solve the domain shift and utilize the overlapping data in multiple hospitals. To address such a challenging scenario, we proposed a new FL framework (Fed-CRFD). Specifically, an intra-client feature disentanglement scheme was proposed to form a unified feature distribution. To eliminate the bias in vertical space, we further proposed a cross-client latent representation consistency constraint to directly narrow the gap between the latent features of the same vertical subject. Our method was assessed on two MRI reconstruction datasets. Experimental results showed that the proposed Fed-CRFD surpassed the state-of-art methods by a large margin. 

\section*{Acknowledgments}
This work was supported by Shenzhen Science and Technology Innovation Committee (NO.~GJHZ20210705141812038), the Key-Area Research and Development Program of Guangdong Province, China (No. 2018B010111001), National Key R\&D Program of China (2018YFC2000702) and the Scientific and Technical Innovation 2030---``New Generation Artificial Intelligence'' Project (No. 2020AAA0104100).


\bibliographystyle{model2-names.bst}\biboptions{authoryear}
\bibliography{medima-template}

\end{document}